\documentclass[10pt, conference, twocolumn]{IEEEtran}
\usepackage{amsfonts}
\usepackage{amssymb}
\usepackage{graphicx}
\usepackage{amsmath}
\usepackage{multirow}
\usepackage{mathtools}
\usepackage{graphicx}
\usepackage{subcaption}
\usepackage{cite}
\usepackage{color}
\usepackage{xcolor}
\usepackage{soul}
\usepackage{makecell}
\usepackage{booktabs}
\usepackage{microtype}
\usepackage{textcomp}
\usepackage{siunitx}
\usepackage{tabu}
\usepackage{float}
\usepackage{tabularx}
\usepackage{listings}
\newcolumntype{C}{ >{\centering\arraybackslash}X}
\usepackage{bm}
\usepackage{threeparttable}
\usepackage{gensymb}
\def\degree{${}^{\circ}$}
\usepackage{algorithmic}
\usepackage{algorithm}

\definecolor{Trypan_blue}{RGB}{58, 12, 163}
\definecolor{grey}{RGB}{245,245,245}
\definecolor{blue}{RGB}{43,159,236}
\definecolor{green}{RGB}{19,158,33}

\usepackage[version=3]{acro}
\acsetup{
	use-id-as-short=true,
	make-links=false,
	link-only-first=true,
}
\makeatletter

\newcommand{\Rmnum}[1]{\expandafter\@slowromancap\romannumeral #1@}

\makeatother
\DeclareAcronym{AI}{
	long={artificial intelligence},
}
\DeclareAcronym{DRL}{
	long={deep reinforcement learning},
}
\DeclareAcronym{LoS}{
	long={line-of-sight},
}
\DeclareAcronym{NLoS}{
	long={non-line-of-sight},
}
\DeclareAcronym{6G}{
	long={the sixth-generation},
}
\DeclareAcronym{SAGIN}{
	long={space-air-ground integrated network},
}
\DeclareAcronym{CV}{
	long={computer vision},
}
\DeclareAcronym{SSL}{
	long={self-supervised learning},
}
\DeclareAcronym{LSTM}{
	long={long short-term memory},
}
\DeclareAcronym{BiLSTM}{
	long={bidirectional long-short term memory},
	short={Bi-LSTM},
}
\DeclareAcronym{CNN}{
	long={convolutional neural network},
}
\DeclareAcronym{GNN}{
	long={graph neural networks},
}
\DeclareAcronym{GCN}{
	long={graph convolutional network},
}
\DeclareAcronym{MLP}{
	long={multi-layer perceptron},
}
\DeclareAcronym{GAN}{
	long={generative adversarial networks},
}
\DeclareAcronym{CSI}{
	long={channel state information},
}
\DeclareAcronym{SVM}{
	long={supported vector machine},
}
\DeclareAcronym{PE}{
	long={positional encoding},
}
\DeclareAcronym{RNN}{
	long={recurrent neural networks},
}
\DeclareAcronym{FC}{
	long={fully connected},
}
\DeclareAcronym{MEC}{
    long={mobile edge computing},
}
\DeclareAcronym{GRU}{
	long={gated recurrent unit},
}
\DeclareAcronym{GPU}{
	long={graphics processing unit},
}
\DeclareAcronym{CUDA}{
	long={compute unified device architecture},
}
\DeclareAcronym{MAPE}{
	long={mean absolute percentage error},
}
\DeclareAcronym{MAE}{
	long={mean absolute error},
}
\DeclareAcronym{MSE}{
	long={mean square error},
}
\DeclareAcronym{NMSE}{
	long={normalized mean square error},
}
\DeclareAcronym{DL}{
	long={deep learning},
}
\DeclareAcronym{GELU}{
    long={gaussian error linear unit},
}
\DeclareAcronym{LEO}{
    long={low Earth orbit},
}
\DeclareAcronym{MAML}{
    long={model-agnostic meta-learning},
}

\DeclareAcronym{LLM}{
    long={large language models}
}
\DeclareAcronym{SSM}{
    long={state space model}
}
\DeclareAcronym{SGP4}{
    long={simplified general perturbations model}
}
\DeclareAcronym{SWaP}{
	long={size, weight and power},
}
\DeclareAcronym{SNR}{
    long={signal-to-noise ratios}
}
\DeclareAcronym{UPA}{
    long={uniform planar array}
}
\DeclareAcronym{AWGN}{
    long={additive white Gaussian noise}
}
\DeclareAcronym{SiLU}{
    long={sigmoid linear unit}
}
\DeclareAcronym{rFFT}{
    long={real fast Fourier transform}
}

\graphicspath{{Figures/}}

\begin{document}
	 \title{Toward Generalized LEO Channel Prediction: A Rapid Adaptation Time-Frequency Mamba Framework}
     
\author{
	\IEEEauthorblockN{
		Yunxin Geng\IEEEauthorrefmark{1}\IEEEauthorrefmark{2}\IEEEauthorrefmark{3}\IEEEauthorrefmark{4}, 
		Yun Zhao\IEEEauthorrefmark{2}\IEEEauthorrefmark{3}\IEEEauthorrefmark{4}, 
        Weiwei Jiang\IEEEauthorrefmark{1}, Shengfeng Wang\IEEEauthorrefmark{2}\IEEEauthorrefmark{3}\IEEEauthorrefmark{4},
        Haotian Liu\IEEEauthorrefmark{1}, 
		and Zhiqing Wei\IEEEauthorrefmark{1}} 
    \IEEEauthorblockA{\IEEEauthorrefmark{1}Beijing University of Posts and Telecommunications, China}
    \IEEEauthorblockA{\IEEEauthorrefmark{2}State Key Laboratory of Satellite Network, China}
	\IEEEauthorblockA{\IEEEauthorrefmark{3}Shanghai Key Laboratory of
Satellite Network, China}\IEEEauthorblockA{\IEEEauthorrefmark{4}Shanghai Satellite
Network Research Institute Co., Ltd., China}
Email: \{gengyx, jww, haotian\_liu, weizhiqing\}@bupt.edu.cn, yunzhao63@163.com, shengfengwang@hotmail.com 
}

\maketitle

\begin{abstract}
Accurate channel state information (CSI) prediction 
is pivotal for enhancing system capacity in low Earth orbit (LEO) satellite-terrestrial communications within the sixth-generation (6G) vision of global ubiquitous connectivity.
However, the extreme mobility of LEO satellites, coupled with significant propagation delays, induces severe channel aging.
Existing deep learning (DL) methods often struggle with non-stationary dynamics or generalize poorly across diverse global geographies.
To address these challenges, we design the Evo-TFMamba, which is an evolutionary framework integrating time-frequency selective Mamba blocks with meta-learning for generalized CSI prediction.
Experimental results demonstrate that the proposed Evo-TFMamba performs adaptive prediction according to the geographic coverage of LEO satellites. Meanwhile, it significantly enhances the model’s generalization ability, particularly when faced with unseen  channel environments in LEO satellite communication.
\end{abstract}


\section{Introduction}
By deeply integrating network resources across space, aerial and ground dimensions, the \ac{SAGIN} is becoming a promising architecture for \ac{6G} systems, where \ac{LEO} satellites play a critical role in providing global coverage and seamless services~\cite{hu2023holographic}. 
The quality of \ac{LEO} satellite communications relies heavily on the accuracy of \ac{CSI} estimation. 
However, the highly dynamic nature of \ac{LEO} satellites and the long signal propagation distance along links lead to channel aging, which poses significant challenges to the accurate prediction of \ac{CSI}.

To ensure reliable satellite-terrestrial communication, extensive researches hve been conducted.
Benefiting from its massive data processing and autonomous learning capabilities, \ac{DL} can flexibly adapt to changes in wireless channels~\cite{jamshed2025non}. 
Several methods obtain accurate downlink \ac{CSI} by extracting spatial features via \ac{CNN} and temporal features through \ac{LSTM}~\cite{ying2024deep,kadambar2023smart}.
Compared with \ac{CNN} and \ac{LSTM}, self-attention mechanism in Transformer can capture global information from both temporal and frequency perspectives, improving system-level performance in \ac{LEO} satellite communication~\cite{yan2025temporal,zhou2024transformer}. 
The Transformer-\ac{LSTM} channel predictor proposed by Li \textit{et al.} remarkably mitigates the effects of channel aging~\cite{li2025leo}.

Although \ac{DL}-driven methods have demonstrated satisfactory performance in terms of prediction accuracy, it still has issues with poor generalization~\cite{li2025csi}. 
Thanks to the development of \ac{LLM} in recent years, existing researches have achieved accurate prediction by utilizing \ac{LLM} to process matrix-form \ac{CSI} data~\cite{he2025sensing,kazemian2026enhancing}.
Moreover, some studies leverage the reasoning and generalization capabilities of pre-trained \ac{LLM} to adapt to the complex radio propagation conditions in \ac{LEO} satellite communication~\cite{chen2025large}. 
However, \ac{LLM}-based approaches require substantial amounts of training samples and computational resources for adequate pre-training, making it difficult to meet on-board real-time requirements as well as \ac{SWaP} limitations.

While existing methods have made progress in terms of \ac{CSI} prediction accuracy and generalization, there is a lack of research on model generalization capabilities across diverse geographical environment,
which is an essential requirement due to the global coverage characteristic of \ac{LEO} satellite constellation.
To this purpose, we propose an evolutionary time-frequency Mamba framework, named Evo-TFMamba, to achieve geography-adaptive \ac{CSI} prediction in \ac{LEO} satellite networks. 
This novel lightweight framework introduces a synergistic dual-path model that integrates a Mamba encoder. The model utilizes selective \ac{SSM} to capture non-stationary channel evolution, with a frequency-domain path employing spectral attention to characterize dominant Doppler peaks.
The model is further enhanced by a environment-aware gated fusion mechanism that dynamically weights temporal and spectral features based on exogenous environmental variables, 
ensuring adaptive reliability across varying link conditions. 
Furthermore, Evo-TFMamba achieves outstanding generalization capabilities with a \ac{MAML} strategy and a unified loss function that integrates \ac{MSE} regression, spectral consistency, and parameter regularization. Moreover, it enables rapid few-shot adaptation to unseen geographical environments.

\section{Preliminaries}
\label{sec:preliminaries}
We introduce the satellite-terrestrial link channel model first, followed by the problem formulation of \ac{CSI} prediction.
\subsection{Satellite-Terrestrial Link Channel Modeling}
\begin{figure}[!t]
    \centering
    \includegraphics[width=0.45\textwidth]{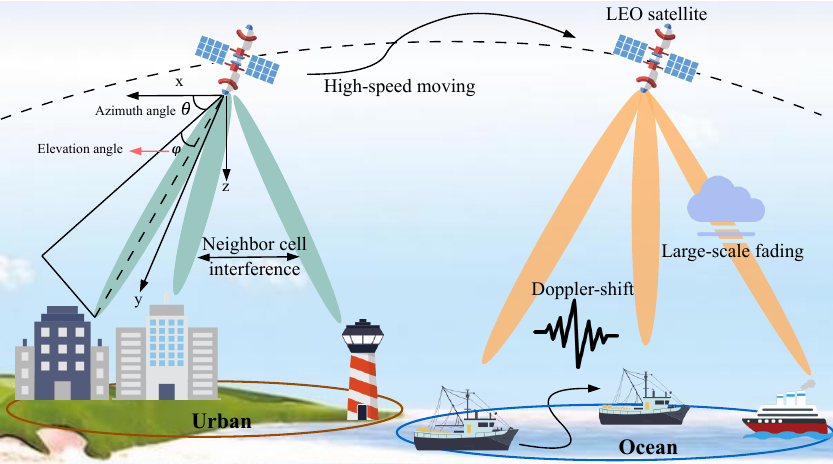}
    \caption{Illustration of the LEO satellite communication.}
    \label{fig:scenario}
\end{figure}

Consider a downlink data transmission scenario, where \ac{LEO} satellite is equipped with a \ac{UPA} comprising $N=N_\text{x} \times N_\text{y}$ antennas to communicate simultaneously with $K$ single-antenna user devices.
The satellite-to-Earth link is characterized by a dominant \ac{LoS} path accompanied by multiple scattered \ac{NLoS} components.
The channel vector at time $t$ is modeled as a Rician fading process~\cite{chen2025large}:
\begin{equation}
\label{eq:H}
    \mathbf{h}[t] = \sqrt{\frac{\mathcal{K}_k}{\mathcal{K}_k + 1}}\;\mathbf{h}_{\text{LoS}}[t] + \sqrt{\frac{1}{\mathcal{K}_k + 1}}\;\mathbf{h}_{\text{NLoS}}[t] \in \mathbb{C}^{N \times 1}
\end{equation}
where $\mathcal{K}_k$ denotes the Rician factor of the link between satellite and the $k^{\textit{th}}$ device. $\mathbf{h}_{\text{LoS}}[t]$ and $\mathbf{h}_{\text{NLoS}}[t]$ represent the \ac{LoS} and \ac{NLoS} components, both normalized to unit average power per antenna.
The \ac{LoS} component incorporates the Doppler-induced phase rotation $\psi_{\text{LOS}}[t] = 2\pi f_d^{\text{full}} (t \Delta t)$ due to relative motion, 
where $f_d^{\text{full}} = \frac{f_c}{c} v  \cos(\alpha_{\text{motion}} - \theta) \cos(\varphi)$ is the full Doppler frequency shift, 
and the spatial steering $\mathbf{s}(\theta) = \left[ 1,  e^{(j\pi\sin\theta)},e^{j2\pi\sin\theta},  \cdots,e^{j\pi(T-1)\sin\theta} \right]^\mathrm{T}$ due to the antenna array geometry.
$f_c$ is the carrier frequency, $c$ is the speed of light, $v$ and $\alpha_{motion}$ are the speed and motion azimuth of terminal, $\theta$ and $\varphi$ are the azimuth and elevation angles of satellite.
Therefore, 
\begin{equation}
\label{eq:H_los}
    \mathbf{h}_{\text{LoS}}[t] = e^{j\,2\pi f_d^{\text{full}}  n\Delta t} \cdot \mathbf{s}(\theta)
\end{equation}
The \ac{NLoS} component arises from multipath scattering due to terrain features, vegetation and atmospheric irregularities. We model it as a superposition of $L$ resolvable paths, i.e.,
\begin{equation}
    \label{eq:H_nlos}
    \mathbf{h}_{\text{NLoS}}[t] = \frac{1}{L}\sum_{\ell=1}^{L} g_\ell  e^{j\,2\pi f_{d,\ell}  t\Delta t + j\phi_\ell} \cdot \mathbf{s}(\theta_\ell)
\end{equation}
where $g_\ell$, $\phi_\ell$, $\theta_\ell$ and $f_{d,\ell}$ represent path loss, random phase, path azimuth angle and path Doppler frequency, respectively.
Additionally, the large-scale channel fading is calculated as $G=10^{-\frac{\text{FSPL}_{\text{dB}} + A_{\text{rain}}}{20}}$, where $\text{FSPL}_{\text{dB}}=20\log_{10}(d_{\text{km}}) + 20\log_{10}\left(\frac{f_c}{10^9}\right) + 92.45$ is free-space path loss~\cite{286098}, and $A_{\text{rain}}$ is the rain attenuation.
Thus the channel vector at time $t$ can be obtained by
\begin{equation}
    \label{eq:H_final}
    \mathbf{h}[t] = G \left(\sqrt{\frac{\mathcal{K}_k}{\mathcal{K}_k+1}}\;\mathbf{h}_{\text{LoS}}[t] + \sqrt{\frac{1}{\mathcal{K}_k+1}}\;\mathbf{h}_{\text{NLoS}}[t]\right) + \mathbf{n}[t]
\end{equation}
where $\mathbf{n}[t]$ is \ac{AWGN}.

\subsection{Channel Prediction Problem Formulation}
$\mathbf{h}[t]$ is converted to a real-valued vector by concatenating the real and imaginary parts of each antenna element 
$\mathbf{x}[t] = \left[\Re\{h_1[t]\}, \cdots, \Re\{h_N[t]\}, \Im\{h_1[t]\}, \cdots, \Im\{h_N[t\} \right]^\mathrm{T}$,
where $\Re\{h_n[t]\}$ and $\Im\{h_n[t]\}$ denote the real and imaginary parts of the element of $\mathbf{h}[t]$, respectively.
To better model the channel state, an auxiliary feature vector $\boldsymbol{\xi} = \left[ d_{\text{km}}, f_d, A_{\text{rain}}, \cos(\varphi), \cos(\alpha_{\text{motion}} - \theta) \right] ^\mathrm{T}$ is used to characterize the propagation scenario, 
where $d_{\text{km}}$ is the satellite-to-terminal distance and $\cos(\alpha_{\text{motion}} - \theta)$ is the directional cosine between the terminal motion azimuth and the satellite azimuth.
Given the historical noisy \ac{CSI} $\mathbf{X} \in \mathbb{R}^{T_P \times 2N}$, the target future \ac{CSI} $\mathbf{Y} \in \mathbb{R}^{T_F \times 2N}$ and the auxiliary feature vector $\boldsymbol{\xi}$, 
where $T_P$ and $T_F$ are the length of past and future horizons,
the objective is to minimize the expected prediction error over the data distribution $\mathcal{D}$:
\begin{equation}
    \label{eq:learning_objective}
    \boldsymbol{\theta}_{\text{mdl}}^\star = \arg\min_{\boldsymbol{\theta}_{\text{mdl}}}\; \mathbb{E}_{(\mathbf{X}, \boldsymbol{\xi}, \mathbf{Y}) \sim \mathcal{D}}\left[\mathcal{L}\!\left(f_{\boldsymbol{\theta}_{\text{mdl}}}(\mathbf{X}, \boldsymbol{\xi}), \mathbf{Y}\right)\right]
\end{equation}
where $\boldsymbol{\theta}_{\text{mdl}}$ denotes all trainable parameters and $\mathcal{L}$ is the loss function. 

\section{System Model}
\label{sec:framework}
In this section, we firstly present our Evo-TFMamba framework, then the \ac{MAML}-based adaptation strategy is introduced.

\subsection{Design of Evo-TFMamba}

\begin{figure*}[!t]
    \centering
    \includegraphics[width=0.90\textwidth]{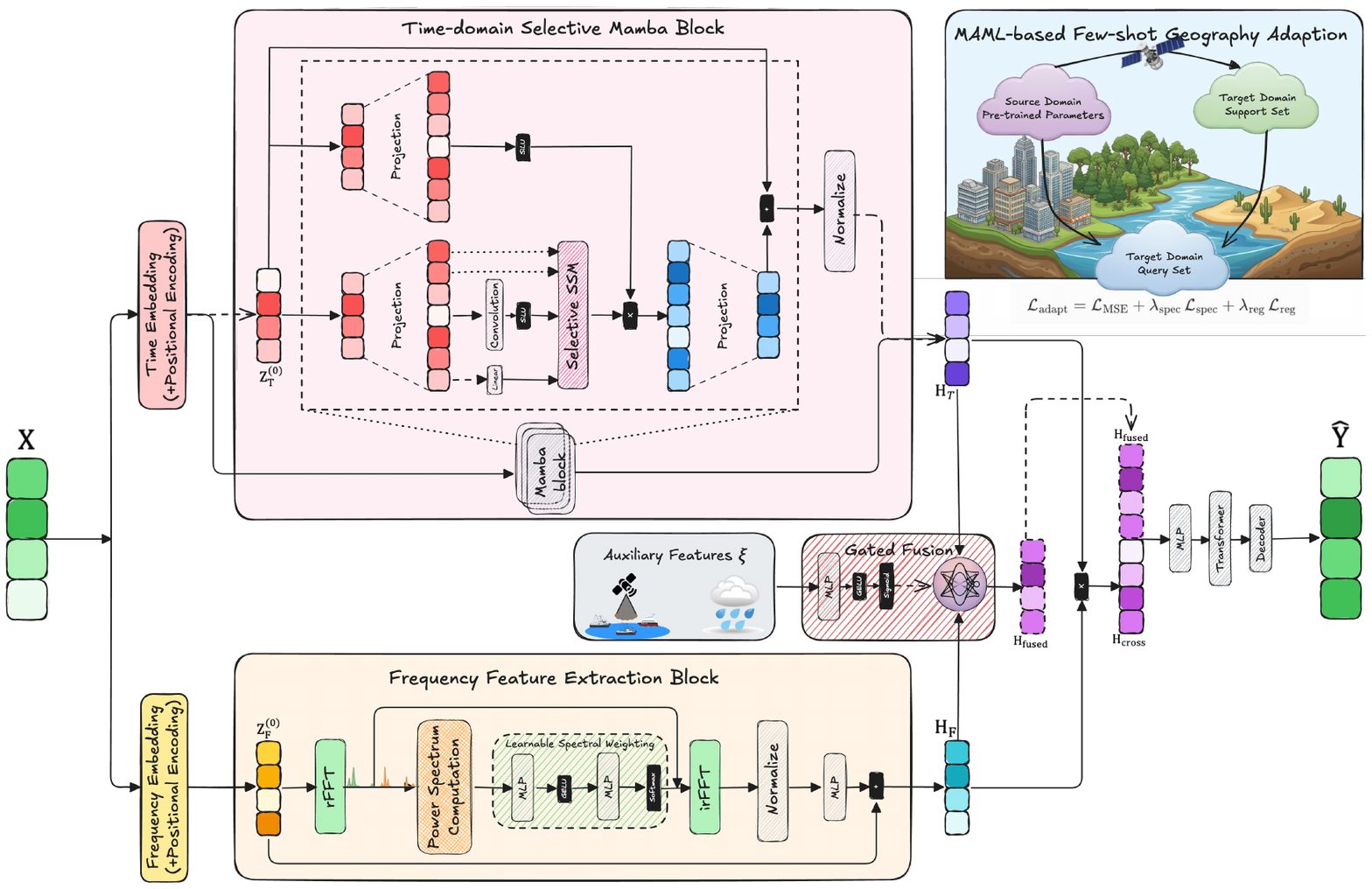}
    \caption{Illustration of the proposed Evo-TFMamba framework.}
    \label{fig:framework}
\end{figure*}

\subsubsection{Normalization and Embedding}
\ac{CSI} features are first normalized using per-feature mean $\mu_i$ and standard deviation $\sigma_i$ computed from the training set:
\begin{equation}
    \label{eq:normalization}
    \tilde{x}_i^{(m)}[t] = \frac{x_i^{(m)}[t] - \mu_i}{\sigma_i + \epsilon}
\end{equation}
where $x_i^{(m)}[t]$ is the $i^{\textit{th}}$ feature of the $t^{\textit{th}}$ time step in the $m^{\textit{th}}$ training sample, and $\epsilon$ is a small constant for numerical stability.
Then the normalized input $\tilde{\mathbf{X}} \in \mathbb{R}^{T_P \times 2N}$ is projected into the model's hidden dimension $D$ and added by a learnable positional encoding in both time-domain and frequency-domain paths, i.e.,
$\mathbf{Z}_T^{(0)} = \tilde{\mathbf{X}}\,\mathbf{W}_e^T + \mathbf{P}_T$ 
and 
$\mathbf{Z}_F^{(0)} = \tilde{\mathbf{X}}\,\mathbf{W}_e^F + \mathbf{P}_F$.

\subsubsection{Time-domain Mamba Encoder}
The time-domain path processes the embedded sequence $\mathbf{Z}_T^{(0)}$ through a stack of $N_{\text{block}}$ selective Mamba blocks. The output of the $n^{\textit{th}}$ block is computed with a pre-norm residual connection:
\begin{equation}
    \label{eq:mambas}
    \mathbf{Z}_T^{(n)} = \mathbf{Z}_T^{(n-1)} + \text{Mamba}_n\!\left(\text{LayerNorm}\!\left(\mathbf{Z}_T^{(n-1)}\right)\right)
\end{equation}
In the full Mamba block implementation, the input $\mathbf{Z} \in \mathbb{R}^{T_P \times D}$ is processed as follows.
It is first projected to an expanded internal dimension $D_{\text{inner}} = D \times \mu$, 
which is then split into two equal halves along the feature dimension $[\mathbf{Z}_1, \mathbf{Z}_2] = \text{chunk}(\mathbf{Z}\,\mathbf{W}_{\text{in}}, 2), \quad \mathbf{Z}_1, \mathbf{Z}_2 \in \mathbb{R}^{T_P \times D_{\text{inner}}/2}$. $\mathbf{Z}_1$ is the branch on which the selective scan operates and $\mathbf{Z}_2$ serves as a gating signal.
Then the selective scan is applied independently to each of the $D_{\text{inner}}/2$ channels of $\mathbf{Z}_1$, producing the scan output
\begin{equation}
    \label{eq:selective_scan}
    \mathbf{o}_{\text{scan}} = \sigma \!\left(\text{LayerNorm}\!\left(\mathbf{Z}_1 \mathbf{W}_s \!\right) \mathbf{W}_g \!\right)\in \mathbb{R}^{T_P \times D_{\text{inner}}/2}
\end{equation}
where $\mathbf{W}_s \in \mathbb{R}^{D_{\text{inner}}/2 \times d_{\text{state}}}$ projects to the state dimension $d_{\text{state}}$ and $\mathbf{W}_g \in \mathbb{R}^{d_{\text{state}} \times D_{\text{inner}}/2}$ projects back to the channel dimension.
$\sigma(\cdot)$ is sigmoid function.
The scan output is combined with the gating branch using a \ac{SiLU} activation:
\begin{equation}
    \label{eq:gated combination}
    \mathbf{o}_{\text{gated}} = \mathbf{o}_{\text{scan}} \odot \text{SiLU}(\mathbf{Z}_2) \in \mathbb{R}^{T_P \times D_{\text{inner}}/2}
\end{equation}
where $\text{SiLU}(x) = x \cdot \sigma(x)$ 
and $\odot$ denotes Hadamard product.
Finally we get 
\begin{equation}
    \label{eq:h_T}
    \mathbf{H}_T = \mathbf{Z}_T^{(0)} + \mathbf{o}_{\text{gated}}\,\mathbf{W}_{\text{out}} \in \mathbb{R}^{T_P \times D} 
\end{equation}
which encodes the temporal evolution of the \ac{CSI} sequence, with the selective state-space mechanism capturing the Doppler-induced phase dynamics.

\subsubsection{Frequency-domain Spectral Attention}
The frequency-domain representation is converted from a real-valued time-domain signal by \ac{rFFT}:
\begin{equation}
    \label{eq:rFFT}
    \mathbf{Z}_F = \text{rFFT}\!\left(\mathbf{Z}_F^{(0)},\; \text{dim} = 1\right) \in \mathbb{C}^{F_{\text{bins}} \times D}
\end{equation}
where $F_{\text{bins}} = \lfloor T_P/2 \rfloor + 1$ is the number of unique frequency bins for a real-valued input of length $T_P$.
Then the spectral energy at each frequency is $P[f, d] = \left|Z_F[f, d]\right|$, where $f=0, \ldots, F_{\text{bins}}-1,\; d=1, \ldots, D$, and $|\cdot|$ denotes the complex modulus.
The mean magnitude across feature channels yields a per-frequency-bin summary $\bar{P}[f] = \frac{1}{D}\sum_{d=1}^{D} P[f, d] \in \mathbb{R}$.
A spectral \ac{MLP} computes attention weights over the frequency bins, which takes the mean power vector $\bar{\mathbf{P}} \in \mathbb{R}^{F_{\text{bins}} \times 1}$ as input:
\begin{equation}
    \label{eq:learnable_weights}
    \mathbf{w} = \mathbf{W}_2  \text{GELU}\!\left(\mathbf{W}_1 \bar{\mathbf{P}} + \mathbf{b}_1\right) + \mathbf{b}_2 \in \mathbb{R}^{F_{\text{bins}} \times 1}
\end{equation}
where $\mathbf{W}_1$, $\mathbf{W}_2$, $\mathbf{b}_1$ and $\mathbf{b}_2$ are the weights and bias of \ac{MLP}, $\text{GELU}(\cdot)$ is the \ac{GELU} activation. 
Let us denote the normalized Fourier coefficients as $\mu[f]$ using the softmax function,
then the attention weights are broadcast across the feature dimension and applied to the complex Fourier coefficients $\widetilde{Z}_F[f, d] = Z_F[f, d] \cdot \mu[f]$. 
The inverse \ac{rFFT} is used to reconstruct the time-domain signal from the attended frequency-domain representation:
\begin{equation}
    \label{eq:irFFT}
    \mathbf{Z}_F^{\text{attn}} = \text{irFFT}\!\left(\widetilde{\mathbf{Z}}_F,\; t = T_P,\; \text{dim} = 1\right) \in \mathbb{R}^{T_P \times D}
\end{equation}
The final frequency-domain representation can be got by
\begin{equation}
    \label{eq:h_F}
    \mathbf{H}_F = \mathbf{Z}_F^{(0)} + \mathbf{W}_{\text{out}}^F \text{LayerNorm}\!\left(\mathbf{Z}_F^{\text{attn}}\right) \in \mathbb{R}^{T_P \times D}
\end{equation}

\addtolength{\topmargin}{0.06in}

\subsubsection{Dynamic Gated Fusion}
The environment-aware gate dynamically balances the contribution of time-domain and frequency-domain features based on the physical scenario parameters encoded in the auxiliary vector $\boldsymbol{\xi}$: 
\begin{equation}
    \label{eq:gate}
    g = \sigma \!\left(\text{GELU}\left(\boldsymbol{\xi}\,\mathbf{W}_{g1} + \mathbf{b}_{g1}\right)\,\mathbf{W}_{g2} + \mathbf{b}_{g2}\!\right)
\end{equation}
The gate is then broadcast across both the time and feature dimensions to match the shape of $\mathbf{H}_T$ and $\mathbf{H}_F$: $\mathbf{G} = g \cdot \mathbf{1}_{T_P \times D} \in \mathbb{R}^{T_P \times D}$, where $\mathbf{1}_{T_P \times D}$ is the all-ones matrix of size $T_P \times D$.
The fused representation is computed as a convex combination:
\begin{equation}
    \label{eq:fusion}
    \mathbf{H}_{\text{fused}} = \mathbf{G} \odot \mathbf{H}_T + (\mathbf{1}_{T_P \times D} - \mathbf{G}) \odot \mathbf{H}_F \in \mathbb{R}^{T_P \times D}
\end{equation}
while $\mathbf{H}_{\text{cross}} = \mathbf{H}_T \odot \mathbf{H}_F$ captures joint time-frequency patterns.
Then $\mathbf{H}_{\text{fused}}$ and $\mathbf{H}_{\text{cross}}$ are concatenated and projected to $D$. 
The result is fed into a single-layer Transformer encoder to further refine the temporal relationships:
\begin{equation}
    \label{eq:transformer}
    \mathbf{H}_{\text{out}} = \text{Transformer}\!\left(\left[\mathbf{H}_{\text{fused}} \;\Vert\; \mathbf{H}_{\text{cross}}\right] \,\mathbf{W}_{\text{cs}} + \mathbf{b}_{\text{cs}} \!\right)
\end{equation}

\subsubsection{Decoder}
The decoder maps the refined representation to the target \ac{CSI} sequence. Temporal mean-pooling is applied to aggregate information across the time dimension, and then the pooled representation is projected to the output dimension:
\begin{equation}
    \label{eq:Y_out}
    \hat{\mathbf{Y}} = \left( \frac{1}{T_P}\sum_{t=0}^{T_P-1} \mathbf{H}_{\text{out}}[t]\right)\,\mathbf{W}_{\text{dec}} + \mathbf{b}_{\text{dec}} \in \mathbb{R}^{T_F \times 2N}
\end{equation}
where $\mathbf{W}_{\text{dec}} \in \mathbb{R}^{D \times (T_F \times 2N)}$ and $\mathbf{b}_{\text{dec}} \in \mathbb{R}^{T_F \times 2N}$ are the decoder weight matrix and bias vector, respectively.

\subsection{\ac{MAML}-based adaptation Strategy}
\begin{algorithm}[t]                    
  \caption{Adaptation for Cross-Environment Prediction}       
  \label{alg:maml_adapt}                  
  \textbf{Initialize}:                    
  \begin{algorithmic}[1]                  
  \STATE Initialize pre-trained parameters $\boldsymbol{\theta}_{\text{mdl}}^{(0)} \leftarrow   \boldsymbol{\theta}_{\text{pretrained}}$.            
  \STATE Prepare $\mathcal{D}_S$ and     
  $\mathcal{D}_Q$ from the target domain. 
  \STATE Set inner-loop learning rate $\gamma$, adaptation steps $K_{\text{adapt}}$, loss weights $\lambda_{\text{spec}}$ and
  $\lambda_{\text{reg}}$.                 
  \end{algorithmic}                                          
  \textbf{Adaptation}:                      
  \begin{algorithmic}[1]                  
  \STATE Initialize $k \leftarrow 0$.                  
  \STATE \textbf{while} $k < K_{\text{adapt}}$ \textbf{do}           
  \STATE \hspace{1em} Zero gradients.
  \STATE \hspace{1em} $\hat{\mathbf{Y}}_S \leftarrow          f_{\boldsymbol{\theta}_{\text{mdl}}^{(k)}}(\mathbf{X}_i^S, \boldsymbol{\xi}_i^S)$ for all $(\mathbf{X}_i^S, \boldsymbol{\xi}_i^S,     
  \mathbf{Y}_i^S) \in \mathcal{D}_S$.
  \STATE \hspace{1em} $\mathcal{L}_{\text{MSE}} \leftarrow \frac{1}{N_S}\sum_{i=1}^{N_S}           
  \|\hat{\mathbf{Y}}_S^{(i)} - \mathbf{Y}_i^S\|_F^2$.  
  \STATE \hspace{1em} Compute $\mathcal{L}_{\text{spec}}$ via rFFT of $\hat{\mathbf{Y}}_S$ and
  $\mathbf{Y}_S$.                         
  \STATE \hspace{1em} $\mathcal{L}_{\text{reg}} \leftarrow \sum_{\ell} \|\boldsymbol{\theta}_{\text{mdl},\,\ell}^{(k)} - \boldsymbol{\theta}_{\text{mdl},\,\ell}^{(0)}\|_2^2$.
  \STATE \hspace{1em} $\mathcal{L}_{\text{adapt}} \leftarrow \mathcal{L}_{\text{MSE}} +\lambda_{\text{spec}}\,\mathcal{L}_{\text{spec}} + \lambda_{\text{reg}}\,\mathcal{L}_{\text{reg}}$.     
  \STATE \hspace{1em}  $\boldsymbol{\theta}_{\text{mdl}}^{(k+1)} \leftarrow 
  \boldsymbol{\theta}_{\text{mdl}}^{(k)} - \alpha \nabla_{\boldsymbol{\theta}_{\text{mdl}}^{(k)}} \mathcal{L}_{\text{adapt}}$.          
  \STATE \hspace{1em}  $\text{NMSE}^{(k+1)} \leftarrow  \text{NMSE}_{\text{dB}}(f_{\boldsymbol{\theta}_{\text{mdl}}^{(k+1)}}, \mathcal{D}_Q)$.                               
  \STATE \hspace{1em} $k \leftarrow k + 1$.      
  \STATE \textbf{end while}               
  \STATE \textbf{return} Adapted model $f_{\boldsymbol{\theta}_{\text{mdl}}^{(K_{\text{adapt}})}}$ 
  \end{algorithmic}                       
  \end{algorithm}
  
Evo-TFMamba is capable of rapidly adapting from the source domain $\mathcal{D}_{\text{src}}$
to the target domain $\mathcal{D}_{\text{tgt}}$
using only a small number of labeled target-domain samples.
Let $\mathcal{D}_S = \{(\mathbf{X}_i^S, \boldsymbol{\xi}_i^S, \mathbf{Y}_i^S)\}_{i=1}^{N_S}$ and $\mathcal{D}_Q = \{(\mathbf{X}_j^Q, \boldsymbol{\xi}_j^Q, \mathbf{Y}_j^Q)\}_{j=1}^{N_Q}$ denote the support set and the query set, respectively.
The adaptation procedure is described in Algorithm~\ref{alg:maml_adapt}.
It is worth noting that a unified adaptation loss function is designed:
\begin{equation}
    \label{eq:adaptation_loss}
    \mathcal{L}_{\text{adapt}} = \mathcal{L}_{\text{MSE}} + \lambda_{\text{spec}}\,\mathcal{L}_{\text{spec}} + \lambda_{\text{reg}}\,\mathcal{L}_{\text{reg}}
\end{equation}
where $\mathcal{L}_{\text{MSE}}$ is the \ac{MSE} regression, $\mathcal{L}_{\text{spec}}$ presents the spectral consistency loss, which constrains the predicted \ac{CSI} to match the target \ac{CSI} in the frequency domain.
$\mathcal{L}_{\text{reg}}$ is the regularization loss that prevents the adapted parameters from deviating too far from the pre-trained initialization.

\section{Experimental Results and Analysis}
\label{sec:experiment}
This section firstly describes the experiments setup and then shows the performance of our proposed Evo-TFMamba.

\subsection{Experimental  Configurations}
\begin{table}[t]
\vspace{0.05in}
		\centering
        \footnotesize
		\caption{Experimental Parameters}
		\label{tab:parameter}
		\begin{tabular}{c|c|c}
			\toprule
			\textbf{Parameter} & \textbf{Value} & \textbf{Description} \\ 
			\midrule
            \multicolumn{3}{c}{\textbf{\ac{LEO} satellite communication parameters}}\\
            \midrule
			$f_c$ & $12~\text{GHz}$ 	& Carrier frequency \\
			\midrule
			$\mathcal{H}$   & $600~\text{km}$	& \ac{LEO} altitude \\
			\midrule
			$\mathcal{V}$ & $7.5~\text{km/s}$ 	& \ac{LEO} satellite velocity\\
			\midrule
			$i$	& $53\degree$ 	& \ac{LEO} inclination\\
			\midrule
			$N$ & 16			& Number of antennas \\
			\midrule
            $K$ & 10			& Number of user devices \\
			\midrule
            $\Delta t$ & $0.5~\text{ms}$			& Time interval between slots \\
			\midrule
			$L$		& 6	& Number of scattering paths \\ 
            \midrule
            ${\tau}_{max}$ & $30~\text{ns}$	& Maximum Delay Spread \\
			\midrule
			$k$	& \makecell[c]{10~dB~(Urban)\\4~dB~(Jungle)}	& Rician factor \\ 
            \midrule
            \multicolumn{3}{c}{\textbf{FreqMamba training parameters}}\\
            \midrule
			$T_P$ & 16		& Number of historical horizons	\\
            \midrule
            $T_F$ & 4		& Number of future horizons	\\
            \midrule
			$D$ & 64		& Dimension of hidden layer	\\
            \midrule
			$N_{block}$ & 2	& Number of Mamba Blocks	\\
            \midrule
			$d_{state}$ & 16 	& State Dimension	\\
            \midrule
			$\mu$ & 2	& Expansion ratio\\
			\bottomrule
		\end{tabular}
        \begin{tablenotes}
        \item $\bullet$ The \ac{LEO} satellite communication parameters are set up based on 3GPP TR 38.821 specifications~\cite{3gpp.38.821}.
        \end{tablenotes}
	\end{table}
The experimental parameters are summarized in Table~\ref{tab:parameter}.
The learning rate of Evo-TFMamba during offline training and \ac{MAML} inner loop are \num{1.0d-4} and \num{1.0d-5}, respectively. 
Training strategies like decay schedule (decay rate $\gamma=0.5$) and early stopping (patience $p=15$) are used to stabilize better convergence and prevent exploding gradients. 
All of the tests were run on an NVIDIA\textsuperscript{TM} GeForce RTX 3080 GPU and the whole program is open sourced on Github
\footnote{[Online]. Available: https://github.com/Xiao-Di/LEO-CSI-Prediction}.

\subsection{Results and Analysis}
\subsubsection{Channel Prediction Performance}
\begin{figure}[tb]
    \centering
    \includegraphics[width=0.40\textwidth]{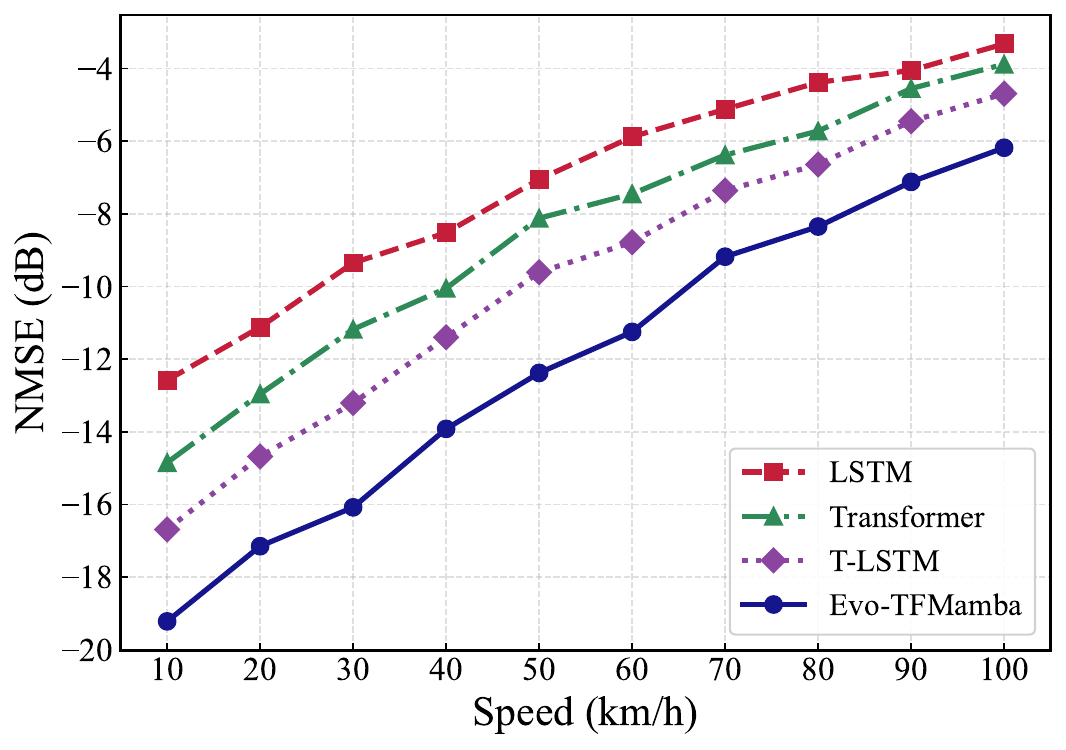}
    \caption{Comparison of different approaches across different device speeds.}
    \label{fig-1:communication result}
\end{figure}
\begin{figure}[tb]
    \centering
    \includegraphics[width=0.40\textwidth]{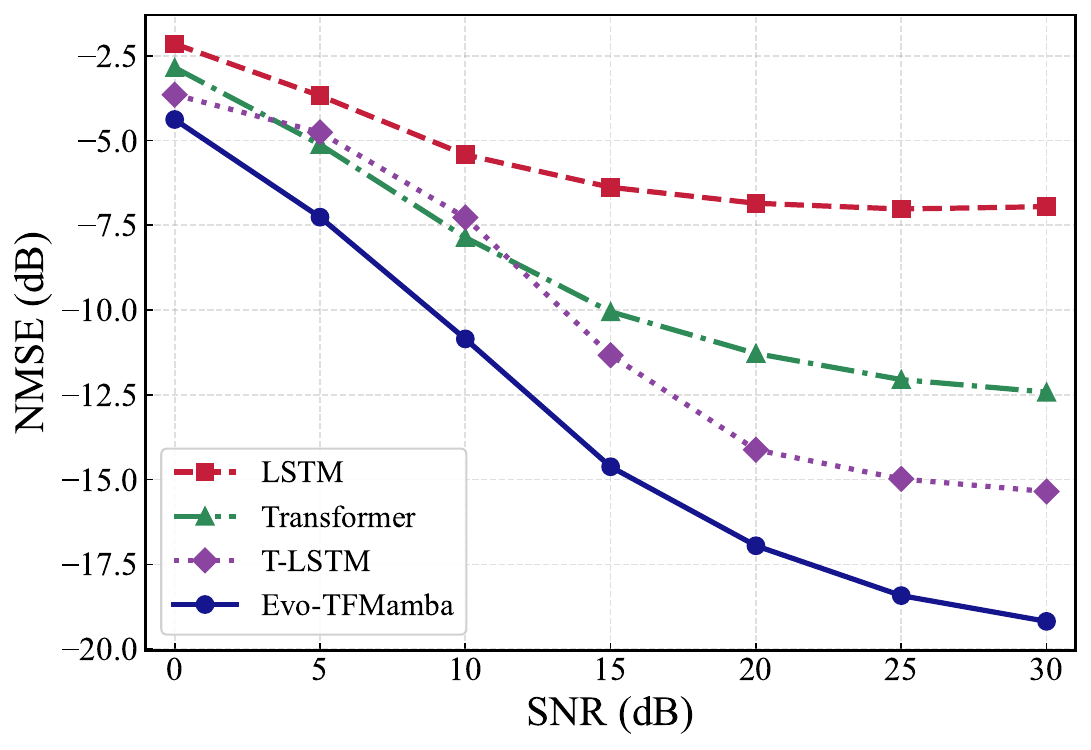}
    \caption{Comparison of different approaches across different \ac{SNR}s.}
    \label{fig-2:communication result}
\end{figure}

We first evaluate the effectiveness of the proposed Evo-TFMamba and existing methods.
It can be observed from Fig.~\ref{fig-1:communication result} that 
Evo-TFMamba outperforms its counterparts across all velocity settings. 
In particular, Evo-TFMamba achieves a \ac{NMSE} reduction of 2.47~dB compared with T-LSTM~\cite{li2025leo} over the speed of 60~km/h.
Fig.~\ref{fig-2:communication result} depicts the comparison result across different \ac{SNR}. 
The prediction accuracy of all schemes improves as \ac{SNR} increases and the proposed Evo-TFMamba is superior to other predictors, especially with high signal quality. 
On average, the proposed Evo-TFMamba reduces \ac{NMSE} by 28\% compared to the best benchmark, i.e., T-LSTM. 

\subsubsection{Cross-Geographic Generalization via \ac{MAML}}
\begin{figure}[tb]
    \centering
    \includegraphics[width=0.40\textwidth]{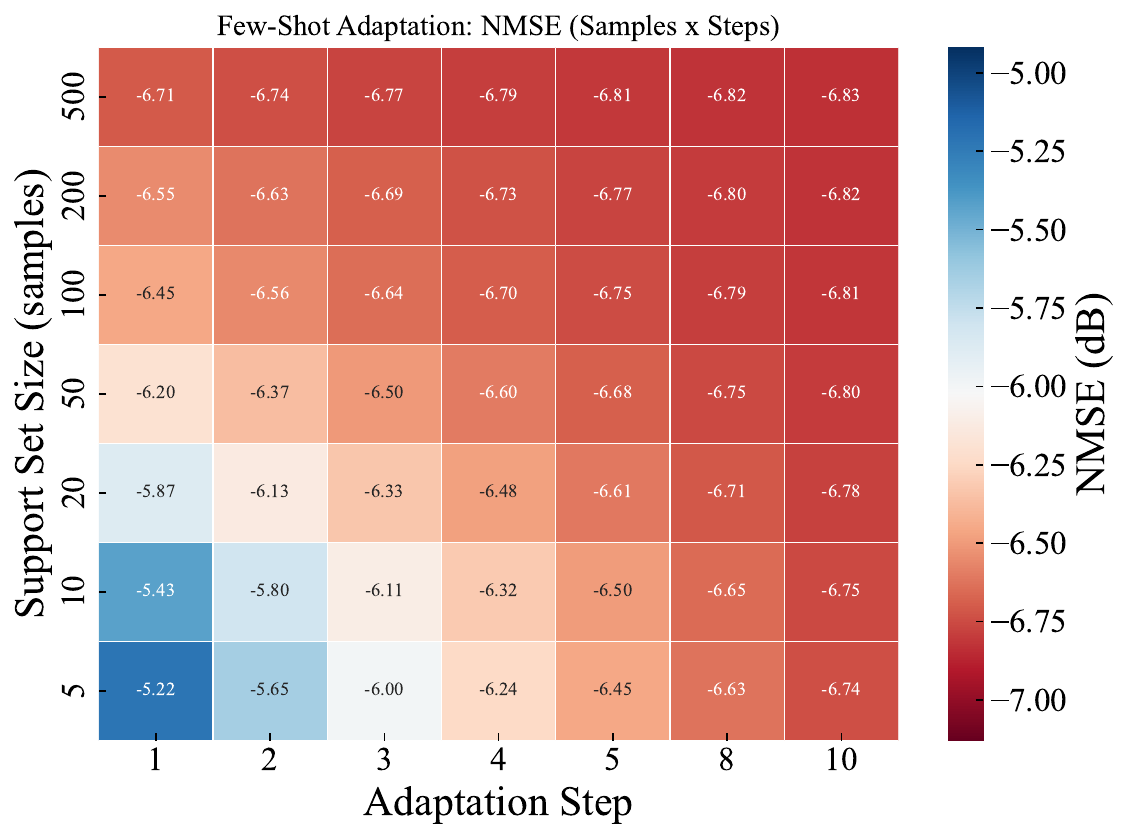}
    \caption{Few-shot adaptation of the proposed Evo-TFMamba.}
    \label{fig:communication result}
\end{figure}
It is assumed that the pre-training dataset is collected from urban environment. As the \ac{LEO} satellite moves, the coverage area shifts from urban areas to rain-forests. We demonstrate the generalization capability of the proposed framework in terms of both adaptation step and support set size.
It can be intuitively observed from the heatmap of Fig.~\ref{fig:communication result} that as the adaptation step or support set size increases, the Evo-TFMamba adapts better to the new geographical environment, as reflected by the decrease in the weighted average \ac{NMSE}.

\begin{figure}[tb]
    \centering        
    \begin{subfigure}{0.40\textwidth}
        \centering
        \includegraphics[width=\textwidth]{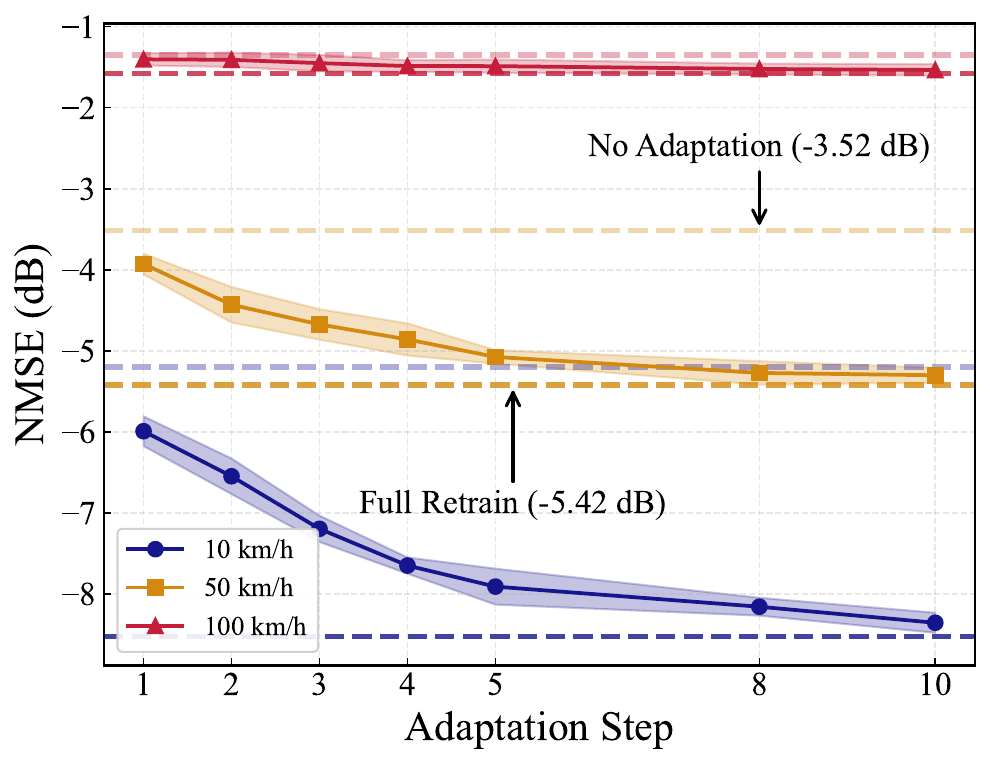}
        \caption{Comparison across different adaptation steps.}
        \label{fig:MAML_a}
    \end{subfigure}
    \begin{subfigure}{0.40\textwidth}
        \centering
        \includegraphics[width=\textwidth]{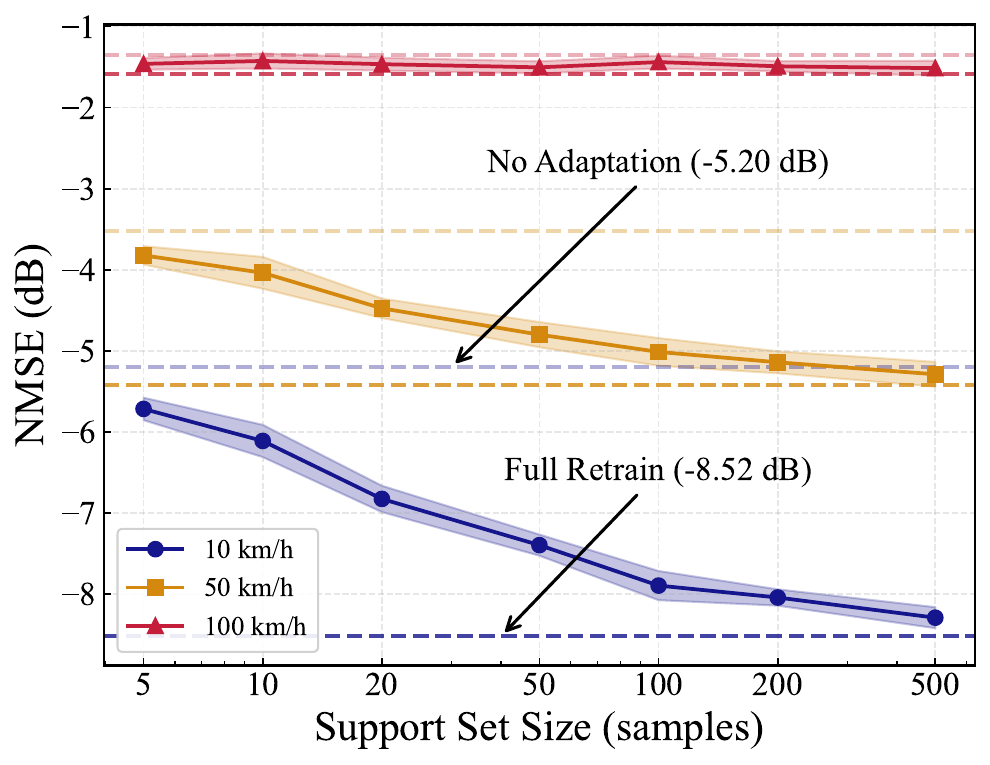}
        \caption{Comparison across different support set sizes.}
        \label{fig:MAML_b}
    \end{subfigure}
    \caption{Illustration of adaptation ability of Evo-TFMamba.} 
    \label{fig:ablation_study_MAML}
\end{figure}

Fig.~\ref{fig:ablation_study_MAML} demonstrates the cross-geographic generalization ability of the proposed Evo-TFMamba in detail, where we validate the effectiveness of the \ac{MAML}-based adaptation strategy across three typical device speeds, i.e., 10~km/h, 50~km/h and 100~km/h. 
The different light-colored and dark-colored horizontal lines respectively represent two baselines of no adaptive transfer and full retraining with various speed settings.
As shown in Fig.~\ref{fig:MAML_a}, when the number of samples in the support set is fixed to 20, once 5 adaptation steps is adopted, the Evo-TFMamba almost achieves the same performance as full retraining. 
We can also conclude from Fig.~\ref{fig:MAML_b} that when the adaptation step is fixed to 5, our proposed model can perform as well as retraining when the number of support sample reaches 100.

\subsubsection{Ablation Study}

\begin{figure}[tb]
    \centering        
    \begin{subfigure}{0.42\textwidth}
        \centering
        \includegraphics[width=\textwidth]{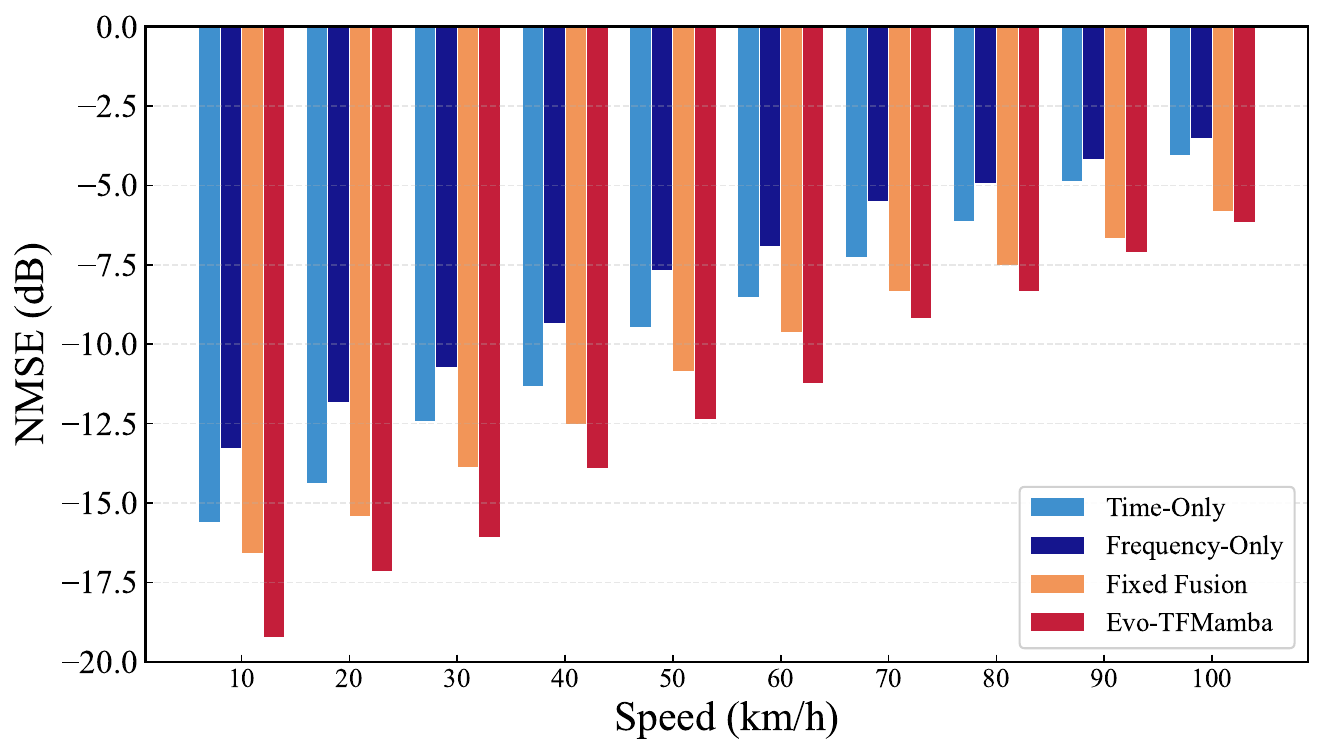}
        \caption{Comparison results across different device speeds.}
        \label{fig:ablation_speed}
    \end{subfigure}
    \begin{subfigure}{0.42\textwidth}
        \centering
        \includegraphics[width=\textwidth]{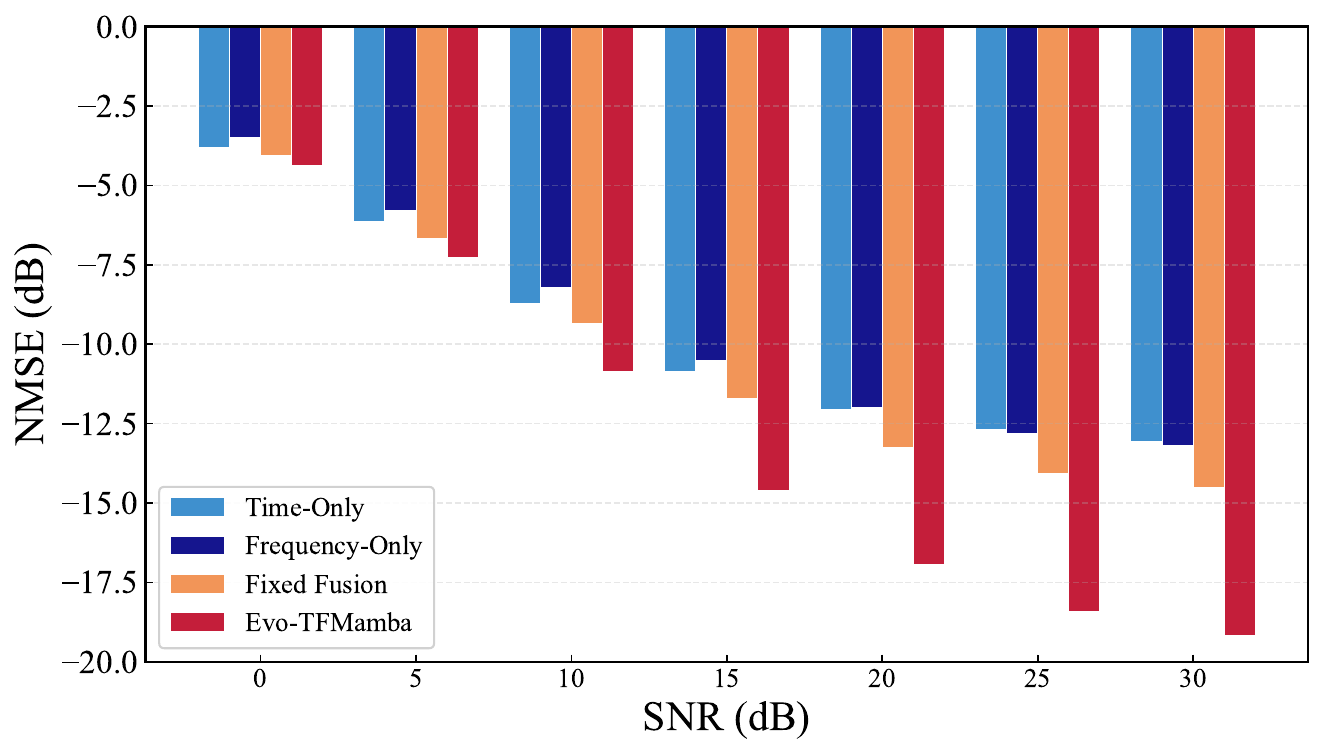}
        \caption{Comparison results across different \ac{SNR}s.}
        \label{fig:ablation_SNR}
    \end{subfigure}
    \caption{Illustration of ablation studies.} 
    \label{fig:ablation_study}
\end{figure}
Sufficient ablation studies are conducted to evaluate the effectiveness of different novel design, e.g. the dual-path time-frequency Mamba structure and the dynamic environment-aware gated fusion.
As illustrated in Fig.~\ref{fig:ablation_study}, on the one hand, we can see from the comparison between \textit{Evo-TFMamba} and \textit{Fixed Fusion} where the time path and the frequency path are equally weighted with 0.5 each, that the environment-aware gated fusion contributes positively to the improvement of the model's performance. 
In particular, the performance advantage of the proposed Evo-TFMamba becomes distinctly evident when \ac{SNR} reaches 30~dB in Fig.~\ref{fig:ablation_SNR}. A significant reduction of about 4~dB can be achieved for our scheme with dynamic gated fusion compared to the \textit{Fixed Fusion}.
On the other hand, by comparing \textit{Evo-TFMamba} with \textit{Time-Only} and \textit{Frequency-Only} respectively, we can observe that the dual-path time-frequency Mamba structure can also significantly enhance the \ac{CSI} prediction performance for satellite-terrestrial links across all the experimental settings of different device speeds and difference \ac{SNR}s.

\section{Conclusion}
\label{sec:conclusion}
This paper proposes Evo-TFMamba, an evolutionary framework for generalized \ac{LEO} \ac{CSI} prediction. It aims to improve accuracy while enhancing adaptation ability across geographical environment.
By integrating a dual-path architecture with environment-aware gated fusion, the model effectively captures the non-stationary time-frequency dynamics of \ac{LEO} channels.
To ensure global robustness, we employ a \ac{MAML} strategy that enables rapid adaptation across diverse geographies.
Experimental results demonstrate that Evo-TFMamba significantly outperforms existing benchmarks, which enables seamless environmental migration within few-shot adaptation steps.
This framework provides a robust and efficient solution for high-dynamic \ac{CSI} prediction in \ac{6G} \ac{SAGIN}.

\section*{Acknowledgement}
This work is supported by National Natural Science Foundation of China under Grant 62401070; Natural Science Foundation of Beijing under Grant L232002.

\bibliographystyle{IEEEtran}
\bibliography{reference}

@article{hu2023holographic,
  title={Holographic beamforming for LEO satellites},
  author={Hu, Xinyuan and Deng, Ruoqi and Di, Boya and Zhang, Hongliang and Song, Lingyang},
  journal={IEEE Communications Letters},
  volume={27},
  number={10},
  pages={2717--2721},
  year={2023},
  publisher={IEEE}
}

@article{yan2025temporal,
  title={Temporal-frequency domain channel prediction for leo satellite communication system: A novel tfformer structure},
  author={Yan, Daifu and Jia, Min and Guo, Qing and Niyato, Dusit},
  journal={IEEE Transactions on Wireless Communications},
  year={2025},
  publisher={IEEE}
}

@article{jamshed2025non,
  title={Non-terrestrial networks for 6G: Integrated, intelligent and ubiquitous connectivity},
  author={Jamshed, Muhammad Ali and Kaushik, Aryan and Dajer, Miguel and Guidotti, Alessandro and Parzysz, Fanny and Lagunas, Eva and Di Renzo, Marco and Chatzinotas, Symeon and Dobre, Octavia A},
  journal={IEEE Communications Standards Magazine},
  year={2025},
  publisher={IEEE}
}

@article{ying2024deep,
  title={Deep learning-based joint channel prediction and multibeam precoding for LEO satellite Internet of Things},
  author={Ying, Ming and Chen, Xiaoming and Qi, Qiao and Gerstacker, Wolfgang},
  journal={IEEE Transactions on Wireless Communications},
  volume={23},
  number={10},
  pages={13946--13960},
  year={2024},
  publisher={IEEE}
}

@inproceedings{kadambar2023smart,
  title={Smart-CSI: Deep learning based low complexity CSI prediction for beyond-5G systems},
  author={Kadambar, Sripada and Chavva, Ashok Kumar Reddy and Lim, Chaiman and Goyal, Ankur and Singh, Divpreet and Kumar, Ashwini and Bal, Samar Ranjan},
  booktitle={2023 IEEE 98th Vehicular Technology Conference (VTC2023-Fall)},
  pages={1--5},
  year={2023},
  organization={IEEE}
}

@article{zhou2024transformer,
  title={Transformer network based channel prediction for CSI feedback enhancement in AI-native air interface},
  author={Zhou, Tao and Liu, Xiangping and Xiang, Zuowei and Zhang, Haitong and Ai, Bo and Liu, Liu and Jing, Xiaorong},
  journal={IEEE Transactions on Wireless Communications},
  volume={23},
  number={9},
  pages={11154--11167},
  year={2024},
  publisher={IEEE}
}

@inproceedings{li2025leo,
  title={LEO Satellite Channel Prediction: A Transformer-LSTM Approach for Outdated CSI},
  author={Li, Wenqing and Omid, Yasaman and Ai, Bo and Niu, Yong and Derakhshani, Mahsa},
  booktitle={2025 IEEE 102nd Vehicular Technology Conference (VTC2025-Fall)},
  pages={1--6},
  year={2025},
  organization={IEEE}
}

@inproceedings{li2025csi,
  title={CSI-ALM: Enhancing Channel State Information Prediction with Semantically Aligned Large Language Models},
  author={Li, Zhaoyang and Yang, Qianqian and Shi, Zhiguo and Xiong, Zehui and Quek, Tony QS},
  booktitle={GLOBECOM 2025-2025 IEEE Global Communications Conference},
  pages={1384--1389},
  year={2025},
  organization={IEEE}
}

@article{he2025sensing,
  title={Sensing-assisted channel prediction in complex wireless environments: An LLM-based approach},
  author={He, Junjie and Ren, Zixiang and Yao, Jianping and Hu, Han and Han, Tony Xiao and Xu, Jie},
  journal={IEEE Wireless Communications Letters},
  year={2025},
  publisher={IEEE}
}

@article{kazemian2026enhancing,
  title={Enhancing 6G Wireless Intelligence: Do LLMs Work for CSI Prediction?},
  author={Kazemian, Mohsen and Jasperneite, J{\"u}rgen},
  journal={arXiv preprint arXiv:2604.04028},
  year={2026}
}

@article{chen2025large,
  title={Large Language Model-Empowered Channel Prediction and Predictive Beamforming for LEO Satellite Communications},
  author={Chen, Zhixiong and Shin, Hyundong and Nallanathan, Arumugam and Chambers, Jonathon},
  journal={arXiv preprint arXiv:2510.10561},
  year={2025}
}

@techreport{3gpp.38.821,
author = {3GPP},
day = {3},
institution = {{3rd Generation Partnership Project (3GPP)}},
month = {4},
note = {Version 16.2.0},
number = {38.821},
title = {{Solutions for NR to support Non-Terrestrial Networks (NTN)}},
type = {Technical report (TR)},
year = {2023}
}

@ARTICLE{286098,
  author={},
  journal={IEEE Std 145-1993}, 
  title={IEEE Standard Definitions of Terms for Antennas}, 
  year={1993},
  volume={},
  number={},
  pages={1-32},
  doi={10.1109/IEEESTD.1993.119664}}

\end{document}